\documentclass[10pt]{article}
\usepackage{amsmath,amssymb, amsthm, amsbsy}
\usepackage[dvips]{graphicx}

\begin{document}

\title{On global in time self-similar solutions of Smoluchowski equation
with multiplicative kernel}
\author{G. Breschi and M. A. Fontelos \\
%EndAName
Instituto de Ciencias Matem\'{a}ticas (ICMAT, CSIC-UAM-UC3M-UCM), \\
C/ Nicol\'{a}s Cabrera 15, 28049 Madrid, Spain.}
\maketitle

\begin{abstract}
We study the similarity solutions (SS) of Smoluchowski coagulation equation
with multiplicative kernel $K(x,y)=(xy)^{s}$ for $s<\frac{1}{2}$. When $s<0$%
, the SS consists of three regions with distinct asymptotic behaviours. The
appropriate matching yields a global description of the solution consisting
of a Gamma distribution tail, an intermediate region described by a
lognormal distribution and a region of very fast decay of the solutions to
zero near the origin. When $s\in \left( 0,\frac{1}{2}\right) $, the SS is
unbounded at the origin. It also presents three regions: a Gamma
distribution tail, an intermediate region of power-like (or Pareto
distribution) decay and the region close to the origin where a singularity
occurs. Finally, full numerical simulations of Smoluchowski equation serve
to verify our theoretical results and show the convergence of solutions to
the selfsimilar regime.
\end{abstract}

\section{Introduction}

Coagulation processes lie at the heart of numerous physical phenomena such
as planetesimal accumulation, mergers in dense clusters of stars, aerosol
coalescence in atmospheric physics, colloids and polymerization and gelation
(see \cite{D72}, \cite{E86}, \cite{JB87}, \cite{L93}). In these processes,
the basic mechanism is the aggregation of two small particles to create
larger particles. Such aggregation will take place with a given probability
that depends on the size of the particles, and the basic issue to solve
concerns the expected evolution of the particle size distribution with time.
The first model for coagulation processes was introduced by Smoluchowski in
1916 (cf. \cite{Smo}). If we denote the particle size distribution by $%
c(x,t) $ and the probabilty of aggretation of two particles of size $x$ and $%
y$ respectively by $K(x,y)$, Smoluchowski equation reads
\begin{equation}
c_{t}\left( x,t\right) =\frac{1}{2}\int_{0}^{x}K\left( x-y,y\right) c\left(
x-y,t\right) c\left( y,t\right) dy-c\left( x,t\right)
\int\limits_{0}^{\infty }K\left( x,y\right) c\left( y,t\right) dy.
\label{smoluchowski}
\end{equation}%
where the first term at the right hand side represents the number of
particles of size $x$ that are created per unit time from the merging of two
particles of sizes $x$ and $x-y$ respectively, and the second term at the
right hand side represents the number of particles of size $x$ that merge
with particles of arbitrary size per unit time.

Despite its formal simplicity, the nonlinear and nonlocal character of
equation (\ref{smoluchowski}) lead to formidable difficulties for the
analysis of its solutions. Explicit solutions are only available for a
limited number of kernels $K(x,y)$ (cf. \cite{Leyvraz} for a general review
and \cite{B1}, \cite{B2} for a broad and recent account of the current
mathematical theory for coagulation-fragmentation models). Two of these
particular cases are $K(x,y)=1$ and $K(x,y)=xy$. Both cases belong to the
broader family of multiplicative kernels $K(x,y)=(xy)^{s}$, $s\in \mathbb{R}$%
. In the first case, $s=0$, solutions exist globally in time while, in the
second, solutions are such that sufficiently high moments $\int
x^{n}c(x,t)dx $ ($n$ large enough) may blow up in finite time giving rise to
a phenomenon known as gelation (see for instance \cite{HDZ}, \cite{ZEH},
\cite{EMP}, \cite{MP}).

In this paper we consider Smoluchowski equation with a multiplicative kernel:%
\begin{equation}
c_{t}(x,t)=\frac{1}{2}\int_{0}^{x}(x-y)^{s}y^{s}c(x-y)c(y)dy-x^{s}c(x)%
\int_{0}^{\infty }y^{s}c(y)dy,  \label{smo}
\end{equation}%
in the case $s<\frac{1}{2}$. In this range of parameters, solutions with all
their moments bounded are expected to exist for all time $t>0$ and behave
asymptotically as $t\rightarrow \infty $ in a selfsimilar manner, that is
\begin{equation}
c(x,t)\sim t^{\alpha }f(t^{\beta }x),  \label{css}
\end{equation}%
in a sense to be precissed and for suitable exponents $\alpha $, $\beta $.
The scaling of equation (\ref{smo}) leads automatically to the relation $%
\alpha =(2s+1)\beta -1$, but $\beta $ remains as a free parameter that needs
to be determined as part of the solution. From the physical point of view, a
result as (\ref{css}) contains the essential information on the behaviour of
the system under consideration and measurable quantities such as exponents
and similarity profiles $f(\xi )$ that can be measured experimentally and
lead to direct physical consequences. It is therefore essential to elucidate
whether such solutions exist and, if so, what is their shape and essential
properties. In a broader sense, equations analogous in structure to (\ref%
{smo}) appear in models of turbulence and results like (\ref{css}) are the
central issue in connection with the development and structure of turbulent
cascades (see \cite{CN1}, \cite{CK1} and references therein). Knowing the
shape and essential properties of similarity solutions $f(\xi )$ is also
relevant in practical applications where a coagulation process takes place
and its evolution is measured experimentally. In these cases, one wants to
know what is the kernel $K(x,y)$ and hence the essential physical processes
involved.

In this paper we compute, by means of matched asymptotic expansions, the
similarity solutions $f(\xi )$ together with the similarity exponent $\beta $
to equation (\ref{smo}) for $s<0$. For $s\in \left( 0,\frac{1}{2}\right) $
we compute the similarity solutions and develop asymptotic expansions for $%
\beta $ as a function of $s$ with $s$ sufficiently small. Finally, full
numerical simulation of (\ref{smo}) is carried out in order to further
support our matched asymptotic expansions and to show convergence of the
solution $c(x,t)$ of (\ref{smo}) towards the selfsimilar regime. Our results
coincide with results obtained by Ca\~{n}izo and Mischler \cite{CM} (see
also \cite{EM}) in the range $s\in \left( -\frac{1}{2},\frac{1}{2}\right) $
concerning asymptotic behaviour of selfsimilar solutions at the origin and
generalize them to other regions in the parameter space as well as provide
further information on the asymptotics away from the origin. In particular,
the case $s\in \left( 0,\frac{1}{2}\right) $ requires a novel procedure
(previously developed and justified with full mathematical rigour in the
context of gelation in finite time in \cite{BF}) for the computation of $%
\beta $ and this translates into special asymptotics for the solution.

\begin{figure}[h]
\centering\includegraphics[width=.95\textwidth]{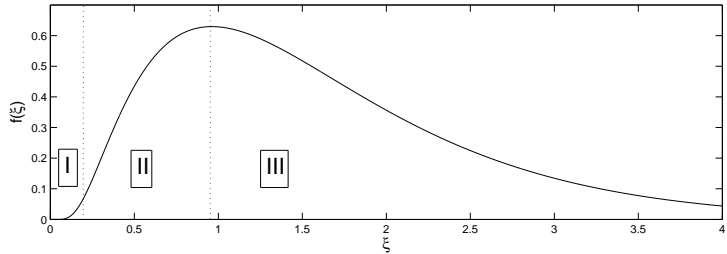}
\caption{{}Structure of the selfsimilar solution for $s<0$. There exist
three regions whose respective behaviours can be described as (I) very vast
decay at the origin, (II) Lognormal distribution function, (III) Gamma
distribution function.}
\label{sketch1}
\end{figure}

A summary of our results is provided in Figures \ref{sketch1} and \ref%
{sketch2}. For $s<0$, $\beta (s)=-1/(1-2s)$ and the similarity solutions
consist of three regions: I) a region of very fast decay to zero near the
origin, II) an intermediate region where the solution approximates a
Lognormal distribution function and III) a region extending to infinity
where the solution approaches a Gamma distribution function. For $s>0$ and
sufficiently small, $\beta (s)=-1-2s+O(s^{2})$ and the solutions also
consist of three regions: I) a singularity developing at the origin, II) a
power-like decay or Pareto distribution function, III) a Gamma distribution
extending up to infinity.
\begin{figure}[h]
\centering\includegraphics[width=.95\textwidth]{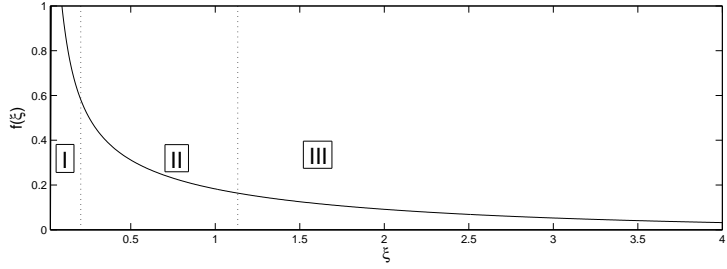}
\caption{{}Structure of the selfsimilar solution for $s\in \left( 0,\frac{1}{%
2}\right) $. There exist three regions whose respective behaviours can be
described as (I) singularity at the origin, (II) Pareto distribution
function, (III) Gamma distribution function.}
\label{sketch2}
\end{figure}

\section{The integrodifferential equation for selfsimilar solutions}

By plugging the selfsimilar expression

\begin{equation*}
c(x,t)=t^{\alpha }f(t^{\beta }x),
\end{equation*}%
into (\ref{smo}), choosing
\begin{equation*}
\alpha =\left( 2s+1\right) \beta -1,
\end{equation*}%
and defining%
\begin{equation*}
\xi :=t^{\beta }x,
\end{equation*}%
we obtain the integrodifferential ordinary differential equation%
\begin{equation}
\left( \left( 2s+1\right) \beta -1\right) f(\xi )+\beta \xi f_{\xi }(\xi )=%
\frac{1}{2}\int_{0}^{\xi }(\xi -\eta )^{s}\eta ^{s}f(\xi -\eta )f(\eta
)d\eta -\xi ^{s}f(\xi )\int_{0}^{\infty }\eta ^{s}f(\eta )d\eta \ .
\label{ss1}
\end{equation}%
$\beta $ is a free parameter that has to be chosen, for a given $s$, from
the condition that all the moments%
\begin{equation*}
M_{n}=\int_{0}^{\infty }x^{n}c(x,t)dx,\ n=1,2,\ldots ,
\end{equation*}%
remain bounded for $0<t<\infty $.

Notice that one can rearrange terms in the more convenient (for the purpose
of analysis) form%
\begin{eqnarray}
&&\left( \left( 2s+1\right) \beta -1\right) f(\xi )+\beta \xi f_{\xi }(\xi )
\notag \\
&=&\frac{1}{2}\int_{0}^{\xi }\eta ^{s}\left[ (\xi -\eta )^{s}f(\xi -\eta
)f(\eta )-2\chi _{\frac{\xi }{2}}(\eta )\xi ^{s}f(\xi )f(\eta )\right] d\eta
\notag \\
&&-\xi ^{s}f(\xi )\int_{\frac{\xi }{2}}^{\infty }\eta ^{s}f(\eta )d\eta ,
\label{ss2}
\end{eqnarray}%
where $\chi _{\frac{\xi }{2}}(\eta )$ is the characteristic function so that
$\chi _{\frac{\xi }{2}}(\eta )=1$ for $\eta \leq \frac{\xi }{2}$ and zero
elsewhere.

A different approach to the problem is through the use of Laplace transform:%
\begin{equation*}
C(\mu ,t)=\int_{0}^{\infty }(e^{-\mu x}-1)c(x,t)dx.
\end{equation*}%
By multiplying equation (\ref{smo}) $(e^{-\mu x}-1)$, integrating in $x$ and
using%
\begin{eqnarray*}
&&\int_{0}^{\infty }e^{-\mu x}\left(
\int_{0}^{x}(x-y)^{s}y^{s}c(x-y,t)c(y,t)dydx\right) \\
&=&\int_{0}^{\infty }\int_{0}^{x}e^{-\mu y}e^{-\mu
(x-y)}(x-y)^{s}y^{s}c(x-y,t)c(y,t)dydx \\
&=&\left( \int_{0}^{\infty }e^{-\mu y}y^{s}c(y,t)dy\right) ^{2},
\end{eqnarray*}%
we arrive at the equation%
\begin{equation*}
C_{t}(\mu ,t)=\frac{1}{2}\left( D_{\mu }^{-s}C(\mu ,t)\right) ^{2},
\end{equation*}%
where%
\begin{equation*}
D_{\mu }^{-s}C(\mu ,t)=\int_{0}^{\infty }(e^{-\mu x}-1)x^{s}c(x,t)dx,
\end{equation*}%
formally represents a $(-s)$-derivative operator. Selfsimilar solutions
would be of the form%
\begin{equation*}
C(\mu ,t)=t^{2s\beta -1}g(t^{-\beta }\mu ),
\end{equation*}%
and satisfy the equation%
\begin{equation}
\left( 2s\beta -1\right) g-\beta \lambda D_{\lambda }g=\frac{1}{2}\left(
D_{\lambda }^{-s}g\right) ^{2},  \label{ss3}
\end{equation}%
where%
\begin{equation*}
\lambda :=t^{-\beta }\mu .
\end{equation*}

If $f(\xi )$ is a solution of (\ref{ss1}), then $\ell ^{1+2s}f\left( \ell
\xi \right) $ is also a solution for any $\ell >0$. Analogously, if $%
g(\lambda )$ is a solution of (\ref{ss3}), then $\ell ^{2s}g\left( \ell
^{-1}\lambda \right) $ is also a solution for any $\ell >0$. For the rest of
this article, when we refer to the selfsimilar solution, we will be
referring to this 1-parameter family (with parameter $\ell $) . For the
purpose of analysis, we will consider a unique representant defined by its
first moment $M_{1}$.

\section{Asymptotic behaviour of selfsimilar solutions}

The particular case $s=0$ with $\beta =-1$ allows direct integration of both
equation (\ref{ss1}) and equation (\ref{ss3}) so that
\begin{equation}
f(\xi )=2e^{-\xi },  \label{sol1}
\end{equation}%
and%
\begin{equation}
g(\lambda )=\frac{-2\lambda }{1+\lambda },  \label{sol2}
\end{equation}%
are their solutions (with first moment given and equal to $2$) respectively.
Of course, (\ref{sol2}) is the Laplace transform of (\ref{sol1})\ as can be
easily verified. If $\beta \neq 1$ then the solution of (\ref{ss3}) is given
by%
\begin{equation*}
g(\lambda )=\frac{-2\lambda ^{-1/\beta }}{1+\lambda ^{-1/\beta }},
\end{equation*}%
and hence%
\begin{equation*}
-\int_{0}^{\infty }e^{-\lambda \xi }\xi f(\xi )d\xi =g^{\prime }(\lambda )=%
\frac{2\beta ^{-1}\lambda ^{-1/\beta -1}}{(1+\lambda ^{-1/\beta })^{2}},
\end{equation*}%
so that, inverting the Laplace transform (see \cite{Abra}), and performing
contour deformation in the complex plane,%
\begin{eqnarray*}
\xi f(\xi ) &=&-\frac{1}{2\pi i}\int_{-i\infty }^{i\infty }e^{\lambda \xi }%
\frac{2\beta ^{-1}\lambda ^{-1/\beta -1}}{(1+\lambda ^{-\beta })^{2}}d\lambda
\\
&=&\frac{1}{2\pi i}\int_{0}^{\infty }e^{-\lambda \xi }\left( \frac{2\beta
^{-1}e^{-i\pi 1/\beta }\lambda ^{-1/\beta -1}}{(1+e^{-i\pi 1/\beta }\lambda
^{-1/\beta })^{2}}-\frac{2\beta ^{-1}e^{i\pi 1/\beta }\lambda ^{-1/\beta -1}%
}{(1+e^{i\pi 1/\beta }\lambda ^{-1/\beta })^{2}}\right) d\lambda .
\end{eqnarray*}%
We find then

\begin{equation}
\xi f(\xi )\sim -\frac{2\beta ^{-1}\sin (\pi /\beta )\Gamma (-1/\beta )}{\pi
}\xi ^{1/\beta },  \label{dec1}
\end{equation}%
as $\xi \rightarrow \infty $, and%
\begin{eqnarray}
\xi f(\xi ) &\sim &\frac{1}{2\pi i}\int_{0}^{\infty }e^{-\lambda \xi }\left(
2\beta ^{-1}e^{i\pi 1/\beta }\lambda ^{1/\beta -1}-2\beta e^{-i\pi 1/\beta
}\lambda ^{1/\beta -1}\right) d\lambda  \notag \\
&=&\frac{2\beta ^{-1}\sin (\pi /\beta )\Gamma (1/\beta )}{\pi }\xi
^{-1/\beta },  \label{dec2}
\end{eqnarray}%
as $\xi \rightarrow 0$. The power-like decay given by (\ref{dec1}) implies
that sufficiently high moments will diverge and therefore solutions with $%
\beta \neq -1$ cannot be allowed. The fact that boundedness of all moments
requires $\beta =-1$ serves to characterize (\ref{sol1}) as a similarity
solution of the second kind in the notation introduced by Barenblatt \cite{B}%
.

For $0<s<\frac{1}{2}$ and $s<0$, explicit integration is not possible and
one has to rely upon perturbation and asymptotic methods in order to study
the solutions.

\subsection{\textit{Case }$0<s<\frac{1}{2}$}

We will follow a methodology identical to the one used in our previous work
\cite{BF} concerning the case $s>\frac{1}{2}$. In that article, we provided
full mathematical proof of formal asymptotics (as $\xi \rightarrow 0$ and $%
\xi \longrightarrow \infty $) analogous to the ones used in the present
work. We start with the asymptotic analysis as $\xi \longrightarrow 0$. By
introducing \ $f(\xi )\sim A\xi ^{\delta }$ into (\ref{ss2}) and letting $%
\xi \rightarrow 0$ we find that the left hand side of (\ref{ss2}) behaves as%
\begin{equation}
\left( \left( 2s+1\right) \beta -1\right) f(\xi )+\beta \xi f_{\xi }(\xi
)\sim \left( \left( 2s+1\right) \beta -1+\delta \beta \right) A\xi ^{\delta
},  \label{lhs1}
\end{equation}%
while the right hand side behaves as%
\begin{eqnarray}
&&\frac{1}{2}\int_{0}^{\xi }(\xi -\eta )^{s}\eta ^{s}\left[ f(\xi -\eta
)f(\eta )-2\chi _{\frac{\xi }{2}}(\eta )f(\xi )f(\eta )\right] d\eta -\xi
^{s}f(\xi )\int_{\frac{\xi }{2}}^{\infty }\eta ^{s}f(\eta )d\eta   \notag \\
&=&A^{2}\int_{0}^{\frac{\xi }{2}}\left[ (\xi -\eta )^{\delta +s}\eta
^{\delta +s}-\xi ^{\delta +s}\eta ^{\delta +s}\right] d\eta +A^{2}\frac{%
2^{-(\delta +s+1)}}{\delta +s+1}\xi ^{2\delta +2s+1}+O(\xi ^{\delta +s})
\notag \\
&=&\left( \int_{0}^{\frac{1}{2}}\left[ \frac{1}{(1-\eta )^{1+s}\eta ^{1+s}}-%
\frac{1}{\eta ^{1+s}}\right] d\eta +\frac{2^{-(\delta +s+1)}}{\delta +s+1}%
\right) A^{2}\xi ^{2\delta +2s+1}+O(\xi ^{\delta +s}),  \label{rhs1}
\end{eqnarray}%
where we have used%
\begin{eqnarray*}
\int_{\frac{\xi }{2}}^{\infty }\eta ^{s}f(\eta )d\eta  &=&-\int_{0}^{\frac{%
\xi }{2}}\eta ^{s}f(\eta )d\eta +\int_{0}^{\infty }\eta ^{s}f(\eta )d\eta  \\
&=&-\frac{2^{-(\delta +s+1)}}{\delta +s+1}A\xi ^{\delta +s+1}+O(1).
\end{eqnarray*}

By comparing (\ref{lhs1}) and (\ref{rhs1}) we conclude \textit{\ }
\begin{equation}
f(\xi )\sim A\xi ^{-1-2s}\ \text{as }\xi \rightarrow 0,  \label{beh1}
\end{equation}%
with%
\begin{equation}
A=\frac{-1}{\int_{0}^{\frac{1}{2}}\left[ (1-\eta )^{-1-s}\eta ^{-1-s}-\eta
^{-1-s}\right] d\eta -\frac{2^{s}}{s}}=\frac{-2\Gamma (-2s)}{\Gamma ^{2}(-s)}%
.  \label{aa}
\end{equation}%
Notice that $A=s+o(s^{3})$ for $s\ll 1$.

On the other hand, for $\xi \gg 1$, by introducing the ansatz $f(\xi )\sim
B\xi ^{\delta }e^{-\xi }$ into (\ref{ss1}) we find that the leading order
contributions from the right and left hand sides are such that%
\begin{equation*}
-\beta BA\xi ^{\delta +1}e^{-B\xi }\sim \frac{B^{2}}{2}\xi ^{2\delta
+1+2s}e^{-B\xi }\int_{0}^{1}(1-\eta )^{s+\delta }\eta ^{s+\delta }d\eta
\end{equation*}%
so that
\begin{equation*}
\delta =-2s\text{,\ }B=-2\beta A\frac{\Gamma (2+2s)}{\Gamma ^{2}(1+s)},
\end{equation*}%
and hence%
\begin{equation}
f(\xi )\sim -2\beta A\frac{\Gamma (2+2s)}{\Gamma ^{2}(1+s)}\xi ^{-2s}e^{-\xi
}\ \text{as }\xi \rightarrow \infty .  \label{beh2}
\end{equation}

As in the case $s=0$, there are also solutions that do not decay
exponentially fast but instead decay algebraically fast. For them, the left
hand. side of (\ref{ss2}) vanishes at leading order, i.e.
\begin{equation}
f(\xi )\sim A\xi ^{-1-2s+1/\beta }\text{ as }\xi \rightarrow \infty .
\label{est}
\end{equation}%
The next order can be computed by plugging (\ref{est}) at the right hand
side of (\ref{ss2}) and solving the resulting equation for the correction $%
\widetilde{f}(\xi )$ to (\ref{est})$:$%
\begin{equation*}
\left( \left( 2s+1\right) \beta -1\right) \widetilde{f}(\xi )+\beta \xi
\widetilde{f}_{\xi }(\xi )\sim A^{2}c_{s,\beta }\xi ^{-1-2s+2/\beta },
\end{equation*}%
where $c_{s,\beta }$ is a numerical constant that can be easily computed.
Hence,%
\begin{equation*}
f(\xi )\sim A\xi ^{-1-2s+1/\beta }+O(\xi ^{-1-2s+2/\beta })\text{ as }\xi
\rightarrow \infty .
\end{equation*}%
Notice that the asymptotics (\ref{est}) agrees with (\ref{dec1}) in the
limit $s\rightarrow 0$. As in (\ref{dec1}), $A$ will be a function of $\beta
$ and it will be the condition that $A$ vanishes (so that (\ref{beh2})
holds) what serves to select the value of $\beta .$

\subsection{\textit{Case }$s<0$}

For $\xi \ll 1$, the last term at the right hand side of (\ref{ss1}) is more
singular than the first term at the left hand side. Hence, by comparing $%
\beta \xi f_{\xi }(\xi )$ with $-\xi ^{s}f(\xi )G$ (where $G$ stands for $%
\int_{0}^{\infty }\eta ^{s}f(\eta )d\eta $ and is assumed to be bounded) we
find a solution with the leading order behaviour $e^{-\frac{G\xi ^{s}}{\beta
s}}$. Since $\beta s>0$ and $s<0$, one expects a very fast decay to zero as $%
\xi \rightarrow 0$ and hence we should neglect the first term at the right
side of (\ref{ss1}). By doing so, we obtain an ordinary differential
equation with solution
\begin{equation}
f(\xi )\sim \frac{A}{\xi ^{2s+1-\frac{1}{\beta }}}e^{-\frac{G\xi ^{s}}{\beta
s}}\ \text{as }\xi \rightarrow 0,  \label{beh3}
\end{equation}%
(see also \cite{Leyvraz} and \cite{CM} where the same behaviour is shown, as
well as the original calculation by \cite{DE}) and, indeed the integral term
is such that%
\begin{equation*}
\int_{0}^{\xi }(\xi -\eta )^{s}\eta ^{s}f(\xi -\eta )f(\eta )d\eta \sim
\int_{0}^{\xi }(\xi -\eta )^{\frac{1}{\beta }-1-s}\eta ^{\frac{1}{\beta }%
-1-s}e^{-\frac{G\eta ^{s}}{\beta s}-\frac{G(\xi -\eta )^{s}}{\beta s}}d\eta
\end{equation*}%
\begin{equation*}
\leq \xi ^{\frac{2}{\beta }-1-2s}e^{-\frac{G\xi ^{s}}{\beta s}2^{1-s}}\ll
\xi ^{-2s-1+\frac{1}{\beta }}e^{-\frac{G\xi ^{s}}{\beta s}}.
\end{equation*}%
Concerning the behaviour as $\xi \rightarrow \infty $, the same argument
that applied for the case $0<s<1$ also applies to the present case and hence
the asymptotics is given by (\ref{beh2}). Note that the asymptotics given by
(\ref{beh3}) and (\ref{beh2}) imply that our assumption that $%
G=\int_{0}^{\infty }\eta ^{s}f(\eta )d\eta $ is bounded is correct.

Notice that the asymptotics given by (\ref{beh1}), (\ref{beh2}) and (\ref%
{beh3}) contain two free parameters: $A$ and $\beta $. The first parameter
can be fixed from the condition that the first moment of $f(\xi )$ (that is,
the total mass) is given and, say, equal to $2$:%
\begin{equation*}
\int_{0}^{\infty }\xi f(\xi )d\xi =2.
\end{equation*}%
The second parameter, the similarity exponent $\beta $, has to be chosen so
that all moments of $f(\xi )$ are bounded. Unfortunately this can only be
done once a global solution to equation (\ref{ss1}) is found. This will be
done, in the next section, explicitly for $s<0$ and by means of a
perturbative approach for $\left\vert s\right\vert \ll 1$.

\section{The selection of the similarity exponent $\protect\beta $}

The similarity exponent $\beta $, which so far is free, can be found in the
case $s<0$ by imposing the condition that all moments of the solution to (%
\ref{ss1}) are bounded. This yields a nonlinear eigenvalue problem that can,
nevertheless, be easily solved based on the asymptotics developed in the
previous section. If we multiply equation (\ref{ss1}) by $\xi $, integrate
by parts the term $\xi ^{2}f_{\xi }$ using the cancellation of boundary
terms (due to the fast decay of $f$ at the origin and infinity) as well as
the relation
\begin{eqnarray*}
&&\frac{1}{2}\int_{0}^{\infty }\int_{0}^{\xi }\xi (\xi -\eta )^{s}\eta
^{s}f(\xi -\eta )f(\eta )d\eta d\xi -\int_{0}^{\infty }\xi ^{s}f(\xi )d\xi
\int_{0}^{\infty }\eta ^{s}f(\eta )d\eta \\
&=&\frac{1}{2}\int_{0}^{\infty }\int_{0}^{\xi }(\xi -\eta )^{s+1}\eta
^{s}f(\xi -\eta )f(\eta )d\eta d\xi \\
&&+\frac{1}{2}\int_{0}^{\infty }\int_{0}^{\xi }(\xi -\eta )^{s}\eta
^{s+1}f(\xi -\eta )f(\eta )d\eta d\xi -\int_{0}^{\infty }\xi ^{s+1}f(\xi
)d\xi \int_{0}^{\infty }\eta ^{s}f(\eta )d\eta \\
&=&\frac{1}{2}M_{1+s}M_{s}+\frac{1}{2}M_{s}M_{1+s}-M_{1+s}M_{s}=0,
\end{eqnarray*}%
we conclude%
\begin{equation*}
\left( \left( 2s-1\right) \beta -1\right) M_{1}=0.
\end{equation*}%
Since $M_{1}>0$, the relation%
\begin{equation}
\beta =-\frac{1}{1-2s},  \label{bw}
\end{equation}%
follows.

The argument above cannot be extended to the case $s\in \left( 0,\frac{1}{2}%
\right) $ where equation (\ref{ss2}) holds. Neither the solution is bounded
at the origin nor cancellations of moments at the right hand side of (\ref%
{ss2}) takes place. We present next the analysis for $s=\varepsilon \ll 1$.
For the purpose of analysis, it will be more convenient to consider the
equation for selfsimilar solutions in the Laplace transform, that is
(equation (\ref{ss3})):%
\begin{equation}
\left( 2\varepsilon \beta -1\right) g-\beta \lambda D_{\lambda }g=\frac{1}{2}%
\left( D_{\lambda }^{-\varepsilon }g\right) ^{2}.  \label{sss}
\end{equation}%
We introduce

\begin{eqnarray*}
\beta &=&-1+B\varepsilon +O(\varepsilon ^{2}), \\
g &=&g_{0}+\varepsilon g_{1}+O(\varepsilon ^{2}),
\end{eqnarray*}%
with%
\begin{equation*}
g_{0}(\lambda )=\frac{-2\lambda }{1+\lambda },
\end{equation*}%
in (\ref{sss}) and obtain%
\begin{eqnarray}
&&\left( -2\varepsilon -1+O(\varepsilon ^{2})\right) \left(
g_{0}+\varepsilon g_{1}+O(\varepsilon ^{2})\right) +\left( 1-B\varepsilon
+O(\varepsilon ^{2})\right) \lambda \left( g_{0}^{\prime }+\varepsilon
g_{1}^{\prime }+O(\varepsilon ^{2})\right)  \notag \\
&=&\frac{1}{2}\left( g_{0}+\varepsilon g_{1}+\varepsilon g_{0,\log
}+O(\varepsilon ^{2})\right) ^{2},  \label{expa}
\end{eqnarray}%
where%
\begin{equation}
g_{0,\log }=2\int_{0}^{\infty }(e^{-\lambda x}-1)\log xe^{-x}dx.
\label{g0log}
\end{equation}%
Since $g_{0}$ satisfies (\ref{sss}) with $\varepsilon =0$, $\beta =-1$, we
obtain by retaining the $O(\varepsilon )$ terms in (\ref{expa}) the equation
\begin{equation*}
-2g_{0}-g_{1}-B\lambda g_{0}^{\prime }+\lambda g_{1}^{\prime
}=g_{0}g_{1}+g_{0}g_{0,\log },
\end{equation*}%
which can be rewritten as%
\begin{equation}
\mathit{L}g_{1}=g_{0}g_{0,\log }+2g_{0}+B\lambda g_{0}^{\prime },  \label{lb}
\end{equation}%
with%
\begin{equation*}
\mathit{L}g_{1}:=-g_{1}+\lambda g_{1}^{\prime }-g_{0}g_{1}.
\end{equation*}%
The question is then: what is the value of $B$ in equation (\ref{lb}) so
that $g_{1}(\lambda )$ is the Laplace transform of a function with all its
moments bounded? In this way, $B$ appears as a compatibility condition for (%
\ref{lb}).

Notice that, for $\left\vert \lambda \right\vert $ sufficiently small, we
can expand $g_{0}(\lambda )$ in the form:
\begin{equation*}
g_{0}=-2\lambda +2\lambda ^{2}+O(\lambda ^{3})
\end{equation*}%
Likewise, $g_{0,\log }(\lambda )$ can be expanded (by standard Taylor
series) as
\begin{eqnarray*}
g_{0,\log }(\lambda ) &=&2\int_{0}^{\infty }(e^{-\lambda x}-1)\log
xe^{-x}dx=-2\frac{\gamma +\log (\lambda +1)}{\lambda +1}+2\gamma \\
&=&2\left( \gamma -1\right) \lambda +\left( 3-2\gamma \right) \lambda
^{2}+O(\lambda ^{3}),
\end{eqnarray*}%
where $\gamma \simeq 0.5772...$ is the Euler's constant (cf. \cite{Abra}).
Hence, the right hand side of (\ref{lb}) can be expanded as%
\begin{eqnarray}
&&g_{0}g_{0,\log }+2g_{0}+B\lambda g_{0}^{\prime }  \notag \\
&=&\left( -2\lambda +2\lambda ^{2}+O(\lambda ^{3})\right) \left( 2+2\left(
\gamma -1\right) \lambda +\left( 3-\gamma \right) \lambda ^{2}+O(\lambda
^{3})\right) +B(-2\lambda +4\lambda ^{2}+O(\lambda ^{3}))  \notag \\
&=&-(2B+4)\lambda +(8-4\gamma +4B)\lambda ^{2}+O(\lambda ^{3})  \label{rhsg0}
\end{eqnarray}

If we look for a solution $g_{1}(\lambda )$ to (\ref{lb}) that is analytic
in a neighborhood of $\lambda =0$, we write%
\begin{equation}
g_{1}=a_{1}\lambda +a_{2}\lambda ^{2}+O(\lambda ^{3}),  \label{ex1}
\end{equation}%
and by straightforward calculation one finds%
\begin{equation}
\mathit{L}g_{1}=-g_{1}+\lambda g_{1}^{\prime }-g_{0}g_{1}=\left(
a_{2}+a_{1}\right) \lambda ^{2}+O(\lambda ^{3}),  \label{rhsg1}
\end{equation}%
so that it is not possible to match the $O(\lambda )$ term in (\ref{rhsg0})
with an equivalent term in (\ref{rhsg1}) unless $B=-2$. Therefore, the
similarity exponent $\beta $ has to be chosen, as a function of $\varepsilon
$, as
\begin{equation*}
\beta (\varepsilon )=-1-2\varepsilon +O(\varepsilon ^{2}).
\end{equation*}%
By comparing the coefficients of $\lambda ^{2}$ in (\ref{rhsg0}) and (\ref%
{rhsg1}) we obtain%
\begin{equation*}
a_{2}+2a_{1}=-4\gamma ,
\end{equation*}%
and provided $a_{1}=0$ (which implies $\int_{0}^{\infty }\xi f_{1}(\xi )d\xi
=0$), one has%
\begin{equation*}
a_{2}=-4\gamma .
\end{equation*}%
By using:%
\begin{eqnarray*}
\int_{0}^{\infty }\left( e^{-\lambda \xi }-1\right) \log \xi e^{-\xi }d\xi
&=&-\frac{\gamma +\log (\lambda +1)}{\lambda +1}+\gamma , \\
\int_{0}^{\infty }\left( e^{-\lambda \xi }-1\right) e^{-\xi }d\xi &=&-\frac{%
\lambda }{\lambda +1},
\end{eqnarray*}%
we can get an explicit expression for $g_{1}(\lambda )$:%
\begin{eqnarray}
g_{1}(\lambda ) &=&-\frac{4\lambda }{(1+\lambda )^{2}}\int_{0}^{\lambda
}\left( \frac{(\gamma +1)z-\log (1+z)}{z}\right) dz  \notag \\
&=&-\frac{4(\gamma +1)\lambda ^{2}}{(1+\lambda )^{2}}-\frac{4\lambda }{%
(1+\lambda )^{2}}Li_{2}(-\lambda ),  \label{g1lam}
\end{eqnarray}%
where $Li_{2}(z)$ is is the dilogarithmic function defined as (see \cite%
{Abra}):%
\begin{equation*}
Li_{2}(z)=-\int_{0}^{z}\frac{\log (1-u)}{u}du,
\end{equation*}%
with the integration contour in the complex plane avoiding the branch-cut
singularity at $\Re (u)<-1$,$\Im (u)=0$.

In the case that $B\neq -2$, the expansion (\ref{ex1}) has to be replaced by%
\begin{equation*}
g_{1}=a_{1}\lambda +a_{2}\lambda \log \lambda +a_{3}\lambda ^{2}+O(\lambda
^{3}),
\end{equation*}%
and by computing the left and right hand sides of (\ref{lb}) we obtain
\begin{equation*}
a_{2}=-(2B+4).
\end{equation*}%
Therefore,%
\begin{equation*}
g(\lambda )=-2\lambda -\varepsilon (2B+4)\lambda \log \lambda +O(\lambda
^{2}),
\end{equation*}%
which is the first order of the expansion in $\varepsilon $ of%
\begin{equation*}
g(\lambda )=-2\lambda ^{1+\varepsilon (B+2)}+O(\lambda ^{2}),
\end{equation*}%
and whose inverse Laplace transform is proportional to $\xi ^{-2-\varepsilon
(B+2)}=\xi ^{-1-2\varepsilon +\frac{1}{\beta }+O(\varepsilon ^{2})}$, in
agreement with (\ref{est}).

\section{Matching at infinity and refined asymptotics at the origin}

In this section we will determine, from the expression for $g_{1}(\lambda )$
given by (\ref{g1lam}), the free coefficients $A$ in the asymptotic
behaviours given by (\ref{beh2}). This will be done for $s=\varepsilon $, $%
\left\vert \varepsilon \right\vert \ll 1$. First, note that by writing $%
\lambda =-1+r$ we can expand (\ref{g1lam}), for $\left\vert r\right\vert \ll
1$, in the form
\begin{equation*}
g_{1}(-1+r)=-\frac{4\left( -Li_{2}(1)+\gamma +1\right) }{r^{2}}+4\frac{\log r%
}{r}+\widetilde{g}(r)
\end{equation*}%
where%
\begin{eqnarray*}
\widetilde{g}(r) &=&O(r^{-1})\text{ as }r\rightarrow 0\text{,} \\
\widetilde{g}(r) &=&-4(\gamma +1)+O\left( \frac{(\log r)^{2}}{r}\right)
\text{ as }r\rightarrow \infty .
\end{eqnarray*}

Observe next that

\begin{equation*}
g_{1}^{\prime }(\lambda )=-\int_{0}^{\infty }e^{-\lambda \xi }\xi f_{1}(\xi
)d\xi .
\end{equation*}%
Hence, inverting the Laplace transform, we get%
\begin{equation}
\xi f_{1}(\xi )=-\frac{1}{2\pi i}\int_{-i\infty }^{i\infty }e^{\lambda \xi
}g_{1}^{\prime }(\lambda )d\lambda  \notag
\end{equation}%
\begin{equation}
=-\frac{e^{-\xi }}{2\pi i}\int_{-i\infty +1}^{i\infty +1}e^{r\xi }\left(
\frac{8\left( -Li_{2}(1)+\gamma +1\right) }{r^{3}}-4\frac{\log r-1}{r^{2}}+%
\widetilde{g}^{\prime }(r)\right) dr,  \label{invlap2}
\end{equation}%
and by using integration contour deformation, the residue theorem and
letting $\xi \rightarrow \infty $, one can easily estimate%
\begin{equation*}
f_{1}(\xi )\sim -4\left( -Li_{2}(1)+\gamma +1\right) \xi e^{-\xi }-4\log \xi
e^{-\xi }+O(e^{-\xi }).
\end{equation*}%
On the other hand, writing $A=1+a\varepsilon $ and expanding (\ref{beh2}) in
$\varepsilon $ we get%
\begin{equation*}
f(\xi )=\allowbreak 2e^{-\xi }-2\varepsilon a\xi e^{-\xi }-4\varepsilon \ln
\xi e^{-\xi }+\allowbreak O\left( \varepsilon e^{-\xi }\right) \ .
\end{equation*}%
Therefore%
\begin{equation*}
a=2\left( -Li_{2}(1)+\gamma +1\right) =2\left( -\frac{\pi ^{2}}{6}+\gamma
+1\right) ,
\end{equation*}%
and then%
\begin{equation}
f(\xi )\sim 2\left( 1+\left( \frac{\pi ^{2}}{3}+2\gamma -2\right)
\varepsilon +O(\varepsilon ^{2})\right) \xi ^{-2\varepsilon +O(\varepsilon
^{2})}e^{-\left( 1+\left( -\frac{\pi ^{2}}{3}+2\gamma +2\right) \varepsilon
+O(\varepsilon ^{2})\right) \xi }\ ,\ \text{as \ }\xi \rightarrow \infty .
\label{decay}
\end{equation}

Next we discuss how to match (\ref{decay}) with the behaviours (\ref{beh1})
(for $\varepsilon >0$) and (\ref{beh3}) (for $\varepsilon <0$) near the
origin. The procedure will yield intermediate regions with distinct features
that we analyse separately.

\subsection{\textit{Case }$\protect\varepsilon >0$}

An explicit solution to (\ref{ss2}) is given by%
\begin{equation*}
f(\xi )=A\xi ^{-1-2\varepsilon },
\end{equation*}%
with $A$ defined in (\ref{aa}). If we introduce a small perturbation $W$ in
the form%
\begin{equation}
f(\xi )=A\frac{1+W}{\xi ^{1+2\varepsilon }},  \label{pert}
\end{equation}%
and linearize equation (\ref{ss2}) for $W$ we deduce%
\begin{eqnarray}
&&-W+\beta \xi W_{\xi }(\xi )=A\int_{0}^{\frac{1}{2}}\left( \frac{W(\xi
(1-\eta ))+W(\xi \eta )}{(1-\eta )^{1+\varepsilon }\eta ^{1+\varepsilon }}-%
\frac{W(\xi \eta )+W(\xi )}{\eta ^{1+\varepsilon }}\right) d\eta  \notag \\
&&-AW(\xi )\xi ^{\varepsilon }\int_{\frac{\xi }{2}}^{\infty }\frac{1}{\eta
^{1+\varepsilon }}d\eta -A\xi ^{\varepsilon }\int_{\frac{\xi }{2}}^{\infty }%
\frac{W(\eta )}{\eta ^{1+\varepsilon }}d\eta .  \label{eqw}
\end{eqnarray}

\bigskip We look for solutions to (\ref{eqw}) in the form%
\begin{equation*}
W=\xi ^{\alpha },
\end{equation*}%
yielding the following equation for $\alpha $:%
\begin{equation}
\frac{1}{\left\vert \beta \right\vert }+\alpha =A\frac{1}{\left\vert \beta
\right\vert }\frac{\Gamma (\alpha -\varepsilon )\Gamma (-\varepsilon )}{%
\Gamma (\alpha -2\varepsilon )},  \label{eqint}
\end{equation}%
with $A$ given by (\ref{aa}). The relation (\ref{eqint}) and this analysis
of small perturbations of (\ref{beh1}) near the origin are not limited to
small values of $\varepsilon $ and is valid if we replace $\varepsilon $ by
an arbitrary $s\in \left( 0,\frac{1}{2}\right) $. Equation (\ref{eqint})
cannot be solved for $\alpha $ in closed form. Nevertheless, if $\varepsilon
$ is small, one can find the solution%
\begin{equation*}
\alpha =\frac{1}{2\left\vert \beta \right\vert }\left( \sqrt{\varepsilon
^{2}\left\vert \beta \right\vert ^{2}+8\varepsilon \left\vert \beta
\right\vert +4}+\varepsilon \left\vert \beta \right\vert -2\right)
=\varepsilon +O(\varepsilon ^{2}).
\end{equation*}%
Notice then that the general form of $W$ is%
\begin{equation}
W(\xi )=C\xi ^{\varepsilon +O(\varepsilon ^{2})}.  \label{w}
\end{equation}%
The constants $C$ and $\beta $ in (\ref{w}) are free and should be chosen so
that the first moment is given (which chooses $C$) and all other moments $%
M_{n}$ ($n>1$) are bounded (that is, the solution decays exponentially fast
at infinity, formula (\ref{decay})). The exact computation of $C$ can only
be done numerically, but we can nevertheless provide a rough sketch the
matching procedure. From (\ref{g1lam}) it is possible to approximate%
\begin{equation}
g_{1}(\lambda )\sim \frac{2(\log \lambda )^{2}}{(\lambda -1)}\text{ as }%
\left\vert \lambda \right\vert \rightarrow \infty ,
\end{equation}%
and by contour deformation we conclude
\begin{eqnarray*}
f_{1}(\xi ) &\sim &\frac{1}{2\pi i}\int_{0}^{\infty }e^{-\xi r}\left( \frac{%
2(\log r+\pi i)^{2}}{1+r}-\frac{2(\log r-\pi i)^{2}}{1+r}\right) dr \\
&=&4\int_{0}^{\infty }e^{-\xi r}\frac{\log r}{1+r}dr\sim -4\log \xi ,\ \text{%
as }\xi \rightarrow 0.
\end{eqnarray*}%
Hence,
\begin{equation}
f_{0}(\xi )+\varepsilon f_{1}(\xi )\sim 2-4\varepsilon \log \xi
+O(\varepsilon ^{2}),  \label{ext}
\end{equation}%
for $\xi \gg e^{-\varepsilon ^{-1}}$. Since $2\xi ^{-2\varepsilon }\sim
2-4\varepsilon \log \xi +O(\varepsilon ^{2})$ for $\xi \gg e^{-\varepsilon
^{-1}}$, we conclude that (\ref{beh1}) and (\ref{ext}) are of the same order
of magnitude for $\xi =O(\varepsilon )$, and this sets the size of the inner
boundary layer where (\ref{beh1}) represents the asymptotic behaviour for
the solution. Between this inner layer and the external region where (\ref%
{decay}) holds, there is an intermediate region where the perturbation $W$
in (\ref{pert}) becomes dominant and therefore $f(\xi )\sim C^{\prime }\xi
^{-1+\alpha -2\varepsilon }$.

\subsubsection{Case $\protect\varepsilon <0$}

In the case $\varepsilon <0$ we obtained the asymptotic term near the origin:

\begin{equation*}
f(\xi )\sim \frac{A}{\xi ^{2\varepsilon +1-\frac{1}{\beta }}}e^{-\frac{G\xi
^{\varepsilon }}{\beta \varepsilon }}\ \text{as }\xi \rightarrow 0.
\end{equation*}%
By expanding%
\begin{equation*}
e^{-\frac{G\xi ^{\varepsilon }}{\beta \varepsilon }}=e^{-\frac{G}{\beta
\varepsilon }}e^{-\frac{G}{\beta }\log \xi }e^{-\frac{G}{2\beta }\varepsilon
(\log \xi )^{2}+...}=e^{-\frac{G}{\beta \varepsilon }}\xi ^{-\frac{G}{\beta }%
}e^{-\frac{G}{2\beta }\varepsilon (\log \xi )^{2}+...},
\end{equation*}%
which is convergent if $\xi \lesssim e^{-\left\vert \varepsilon \right\vert
^{-\frac{1}{2}}}$, and defining%
\begin{equation*}
\ A=e^{\frac{G}{\beta \varepsilon }}a(\varepsilon ),
\end{equation*}%
we conclude $f(\xi )=aO(1)$ and hence%
\begin{eqnarray*}
Q(\xi ) &=&\int_{0}^{\xi }(\xi -\eta )^{\varepsilon }\eta ^{\varepsilon
}f(\xi -\eta )f(\eta )d\eta \leq A^{2}\xi ^{\frac{2}{\beta }-1-2\varepsilon
}e^{-\frac{G\xi ^{\varepsilon }}{\beta \varepsilon }2^{1-\varepsilon }} \\
&\simeq &a^{2}\xi ^{\frac{2}{\beta }-1-2\varepsilon }\left( \xi /2\right) ^{-%
\frac{2G}{\beta }}\simeq \frac{a^{2}}{2^{4}}\xi ^{1+O(\varepsilon )},
\end{eqnarray*}%
for $\xi \lesssim e^{-\left\vert \varepsilon \right\vert ^{-\frac{1}{2}}}$.
By integrating the equation for selfsimilar solution we arrive at the
formula
\begin{equation}
f(\xi )=\frac{e^{\frac{G}{\beta \varepsilon }}}{\xi ^{2\varepsilon +1-\frac{1%
}{\beta }}}e^{-\frac{G\xi ^{\varepsilon }}{\beta \varepsilon }}\left[
a(\varepsilon )+\int_{0}^{\xi }\frac{\eta ^{2\varepsilon +1-\frac{1}{\beta }}%
}{2\beta }e^{\frac{G(\eta ^{\varepsilon }-1)}{\beta \varepsilon }}Q(\eta
)d\eta \right] .  \label{inteq}
\end{equation}%
Given the asymptotics for $Q(\xi )$ we find that the integral at the right
hand side of (\ref{inteq}) is $O(e^{-\left\vert \varepsilon \right\vert ^{-%
\frac{1}{2}}})$ for $\xi \lesssim e^{-\left\vert \varepsilon \right\vert ^{-%
\frac{1}{2}}}$. Hence, we can neglect the contribution to the integral from
the region $\xi \lesssim e^{-\left\vert \varepsilon \right\vert ^{-\frac{1}{2%
}}}$ and integrate outside this region using (\ref{decay}) so that%
\begin{equation*}
Q(\xi )\simeq 4(1+2\left( \frac{\pi ^{2}}{3}+2\gamma -2\right) \varepsilon
+O(\varepsilon ^{2}))e^{-\left( 1+\left( -\frac{\pi ^{2}}{3}+2\gamma
+2\right) \varepsilon +O(\varepsilon ^{2})\right) \xi }\int_{0}^{\xi }(\xi
-\eta )^{-\varepsilon }\eta ^{-\varepsilon }d\eta .
\end{equation*}%
Since%
\begin{equation*}
\int_{0}^{\xi }(\xi -\eta )^{-\varepsilon }\eta ^{-\varepsilon }d\eta
=(1+2\varepsilon +O(\varepsilon ^{2}))\xi ^{1-2\varepsilon },
\end{equation*}%
we will have a solution to (\ref{inteq}) provided
\begin{equation*}
a(\varepsilon )=-\int_{0}^{\infty }\frac{\eta ^{2\varepsilon -\frac{1}{\beta
}}}{2\beta }e^{\frac{G\eta ^{\varepsilon }}{\beta \varepsilon }}Q(\eta
)d\eta \simeq -\int_{0}^{\infty }\frac{\eta ^{2\varepsilon -\frac{1}{\beta }+%
\frac{G}{\beta }}}{2\beta }Q(\eta )d\eta
\end{equation*}%
\begin{equation*}
=-\frac{2}{\beta }(1+2\left( \frac{\pi ^{2}}{3}+2\gamma -1\right)
\varepsilon +O(\varepsilon ^{2}))\int_{0}^{\infty }\eta ^{1-\frac{1}{\beta }+%
\frac{G}{\beta }}e^{-\left( 1+\left( -\frac{\pi ^{2}}{3}+2\gamma +2\right)
\varepsilon +O(\varepsilon ^{2})\right) \eta }d\eta
\end{equation*}%
\begin{equation*}
=-\frac{2}{\beta }\frac{1+2\left( \frac{\pi ^{2}}{3}+2\gamma -1\right)
\varepsilon +O(\varepsilon ^{2})}{1+\left( -\frac{\pi ^{2}}{3}+2\gamma
+2\right) \varepsilon +O(\varepsilon ^{2})}\Gamma (2-\frac{1}{\beta }+\frac{G%
}{\beta }),
\end{equation*}%
and using%
\begin{equation}
\beta =-1-2\varepsilon +O(\varepsilon ^{2}),  \label{d}
\end{equation}%
we find%
\begin{eqnarray}
G &=&2\frac{\left( 1+\left( \frac{\pi ^{2}}{3}+2\gamma -2\right) \varepsilon
+O(\varepsilon ^{2})\right) }{\left( 1+\left( -\frac{\pi ^{2}}{3}+2\gamma
+2\right) \varepsilon +O(\varepsilon ^{2})\right) }\Gamma (1-\varepsilon )
\label{e} \\
&=&\allowbreak 2+\left( 2\gamma +\frac{4}{3}\pi ^{2}-8\right) \varepsilon
+O\left( \varepsilon ^{2}\right) ),  \notag
\end{eqnarray}%
$\allowbreak $providing the value of the free parameter $G$ for the
asymptotic value of the solution at the origin. Hence, the matching is now
complete and all parameters determined for the selfsimilar solution $f(\xi )$%
. We can also find%
\begin{equation}
a(\varepsilon )=\allowbreak 2+\left( 4\gamma ^{2}-16\gamma +\frac{8}{3}%
\gamma \pi ^{2}+2\pi ^{2}-12\right) \varepsilon +O\left( \varepsilon
^{2}\right) .  \label{f}
\end{equation}%
Finally, by expanding the first factor at the right hand side of (\ref{inteq}%
) and using (\ref{d}),\ (\ref{e}) and (\ref{f}) we conclude
\begin{equation}
\frac{e^{\frac{G}{\beta \varepsilon }}}{\xi ^{2\varepsilon +1-\frac{1}{\beta
}}}e^{-\frac{G\xi ^{\varepsilon }}{\beta \varepsilon }}\sim \frac{1}{\xi
^{-\left( 2\gamma +\frac{4}{3}\pi ^{2}-4\right) \varepsilon }}e^{\varepsilon
(\log \xi )^{2}},  \label{apx}
\end{equation}%
for any $e^{-\left\vert \varepsilon \right\vert ^{-1}}\ll \xi \ll
e^{-\left\vert \varepsilon \right\vert ^{-1/2}}$. Notice that we can rewrite
the right hand side of (\ref{apx}) as $e^{\varepsilon (\log \xi
)^{2}+\varepsilon \left( 2\gamma +\frac{4}{3}\pi ^{2}-4\right) \log \xi }$,
which is a function of $\log \xi $ that decays at $\pm \infty $ and whose
maximum value is $e^{-\varepsilon \left( \gamma +\frac{2}{3}\pi
^{2}-2\right) ^{2}}=1+O(\varepsilon )$ (as one can easily verify). For $\xi
\lesssim e^{-\left\vert \varepsilon \right\vert ^{-\frac{1}{2}}}$, the
integral at the right hand side of (\ref{inteq}) is still negligible, while (%
\ref{apx}) is $1+O(\varepsilon )$. $\ $At some $\xi >e^{-\left\vert
\varepsilon \right\vert ^{-\frac{1}{2}}}$, $f(\xi )$ reaches its maximum and
starts to decrease due to the increase of the integral at the right hand
side of (\ref{inteq}) and eventually decays exponentially fast as given by (%
\ref{decay}). Hence, we can distinguish three regions: a) the region $\xi
\lesssim e^{-\left\vert \varepsilon \right\vert ^{-1}}$ where%
\begin{equation*}
f(\xi )\sim a(\varepsilon )e^{\frac{2+O(\varepsilon )}{\left\vert
\varepsilon \right\vert }}e^{-\frac{2+O(\varepsilon )}{\left\vert
\varepsilon \right\vert \xi ^{\left\vert \varepsilon \right\vert }}%
-(2+O(\varepsilon ))\log \xi },
\end{equation*}%
which decays extremely fast to zero as $\xi \rightarrow 0$ (faster than any
power), b) the region $e^{-\left\vert \varepsilon \right\vert ^{-1}}\ll \xi
\ll e^{-\left\vert \varepsilon \right\vert ^{-\frac{1}{2}}}$ where
\begin{equation}
f(\xi )\sim a(\varepsilon )e^{\varepsilon (\log \xi )^{2}+\varepsilon \left(
2\gamma +\frac{4}{3}\pi ^{2}-4\right) \log \xi },  \label{logn}
\end{equation}%
and where a transition between the first region and the maximum value of $%
f(\xi )$ takes place, and c) outer region $\xi \gg e^{-\left\vert
\varepsilon \right\vert ^{-\frac{1}{2}}}$ where $f(\xi )$ is a small
perturbation of $2e^{-\xi }$ for $\xi \lesssim e^{\left\vert \varepsilon
\right\vert ^{-1}}$ and the asymptotic behaviour is given by (\ref{decay}).

It is worth noting that the behaviour implied by (\ref{logn}) is similar to
that of a lognormal distribution, while the asymptotics (\ref{decay})
corresponds to a gamma distribution.

\section{Selfsimilar solutions for $\protect\varepsilon \allowbreak =-n$}

In the particular case when the kernel is of the form $K(x,y)=(xy)^{-n}$,\ ($%
n=1,2,...$), equation (\ref{ss2}), written in terms of $F(\xi )=f(\xi )/\xi
^{n}$, takes the form%
\begin{equation}
\left( \left( -n+1\right) \beta -1\right) \xi ^{n}F+\beta \xi ^{n+1}F_{\xi
}(\xi )=\frac{1}{2}\int_{0}^{\xi }F(\xi -\eta )F(\eta )d\eta -F(\xi
)\int_{0}^{\infty }F(\eta )d\eta .  \label{ssn}
\end{equation}%
By defining the Laplace transform%
\begin{equation*}
G(\lambda )=\int_{0}^{\infty }e^{-\lambda \xi }F(\xi )d\xi ,
\end{equation*}%
equation (\ref{ssn}) takes the form%
\begin{equation}
(-1)^{n}\left( \left( -n+1\right) \beta -1\right) \frac{d^{n}G(\lambda )}{%
d\lambda ^{n}}+(-1)^{n+1}\beta \frac{d^{n+1}}{d\lambda ^{n+1}}\left( \lambda
G(\lambda )\right) =\frac{1}{2}G^{2}(\lambda )-G(0)G(\lambda ).  \label{ssnl}
\end{equation}%
By suitably rescaling variable and function, we can assume $G(0)=1$. If we
seek for a solution that is analytic near the origin $\lambda =0$,
\begin{equation*}
G(\lambda )=1+\sum_{m=1}^{\infty }a_{m}\lambda ^{m},
\end{equation*}%
we find that the right hand side of (\ref{ssnl}) is%
\begin{equation*}
RHS=-\frac{1}{2}+\frac{a_{1}^{2}}{2}\lambda ^{2}+O(\lambda ^{3}).
\end{equation*}%
That is, there is no $O(\lambda )$ term. Hence, the linear right hand side
of (\ref{ssnl}) cannot contain $O(\lambda )$ term. This implies $a_{n+1}=0$
or%
\begin{equation}
\beta =\beta _{n}=-\frac{1}{2n+1}.  \label{betan}
\end{equation}%
The first possibility ($a_{n+1}=0$) would imply that the $M_{n+1}$ moment
vanishes, which is not possible for a positive solution. Therefore, the
similarity exponent will generically be given by (\ref{betan}) as we know
from (\ref{bw}) and will verify numerically in the next section.

Finally, notice the possibility of a pole of $G(\lambda )$ at $\lambda =-a$ (%
$a$ real and positive) which is a local solution to (\ref{ssnl}) where the
dominant contributions balance:%
\begin{equation}
(-1)^{n+1}\beta _{n}\frac{d^{n+1}}{d\lambda ^{n+1}}\left( \lambda G(\lambda
)\right) \simeq \frac{1}{2}G^{2}(\lambda ).  \label{dbal}
\end{equation}%
By inserting $G(\lambda )=c/(\lambda +a)^{\alpha }$ into (\ref{dbal}) we
find, at leading order, $\alpha =-(n+1)$, $c=2a\beta _{n}\frac{(2n+1)!}{n!}$
and therefore
\begin{equation}
G(\lambda )\sim \frac{2(2n)!}{n!}\frac{a}{(\lambda +a)^{n+1}},\ \text{as }%
\lambda \rightarrow -a.  \label{asg}
\end{equation}%
This implies a generic behaviour of $f(\xi )$ (inverting Laplace transform)
of the form%
\begin{equation*}
f(\xi )\sim C_{n}a\xi ^{n}e^{-a\xi }\text{ as }\xi \rightarrow \infty ,
\end{equation*}%
where $C_{n}$ can be computed straightforwardly by evaluation of the residue
given by the pole of $G(\lambda )$ at $\lambda =-a$ when inverting the
Laplace transform. The parameter $a$ is free, but should be estimated from
the condition $G(0)=1$ once $G(\lambda )$ is evaluated in $\Re\lambda <0$. \
This selection of the free parameter $a$ can be done analytically for $n\gg
1 $. In this case, by evaluating the right hand side of (\ref{asg}) at $%
\lambda =0$ we find the identity%
\begin{equation*}
\frac{(2n)!}{n!}a^{-n}=1,
\end{equation*}%
which yields, using Stirling's formula,%
\begin{equation*}
a=\left( \frac{(2n)!}{n!}\right) ^{\frac{1}{n}}\simeq \left( \frac{%
e^{-2n}(2n)^{2n}\sqrt{4\pi n}}{e^{-n}n^{n}\sqrt{2\pi n}}\right) ^{\frac{1}{n}%
}\simeq 4e^{-1}n.
\end{equation*}%
Since%
\begin{equation*}
C_{n}\sim \frac{2(2n)!}{(n!)^{2}}\sim \frac{2e^{-2n}(2n)^{2n}\sqrt{4\pi n}}{%
e^{-2n}n^{2n}(2\pi n)}=2^{2n+1}\frac{1}{\sqrt{\pi n}},
\end{equation*}%
we can conclude%
\begin{equation}
f(\xi )\sim \frac{8e^{-1}}{\sqrt{\pi }}\sqrt{n}4^{n}\xi ^{n}e^{-4e^{-1}n\xi
}\ \text{as }\xi \rightarrow \infty ,  \label{asn}
\end{equation}%
If we take the right hand side of (\ref{asn}) as valid for any $\xi >0$,
then we find a local behaviour near the $n$-dependent maximum of $f(\xi )$
described by
\begin{equation}
f(\xi )\sim \frac{8e^{-1}}{\sqrt{\pi }}\sqrt{n}e^{-\frac{1}{2}n(\xi
-e/4)^{2}}=\sqrt{n}\Phi \left( \sqrt{n}(\xi -\xi _{0})\right) ,  \label{ssm}
\end{equation}%
with $\xi _{0}=e/4$, and $\Phi $ a gaussian function. Hence, $f(\xi )$ would
approach a Dirac delta as $n\rightarrow \infty $. Of course, the assumption
that (\ref{asn}) is valid for any $\xi >0$ is not correct, but the
conclusion that $f(\xi )$ converges to a certain rescaled (with $n$)
function $\Phi $ as $n\rightarrow \infty $ will be verified numerically in
the next section.

\section{\protect\bigskip Numerical computation of selfsimilar solutions}

Equation (\ref{inteq}), which is valid for $s<0$, provides a simple way to
numerically compute the selfsimilar solutions. Notice first that the term $%
Q(\eta )$ involves an integral over the interval $\left[ 0,\xi /2\right] $.
Hence, all information at the right hand side of (\ref{inteq}) concerning
values of $f(\eta )$ for $\eta >\xi $ is limited to the real parameter $%
G=\int_{0}^{\infty }\eta ^{s}f(\eta )d\eta $. We will take an arbitrary
value of $G$ (remember that the selfsimilar solutions, for a given $s,$ are
indeed a 1-parameter family $\ell ^{1+2s}f\left( \ell \xi \right) $ so that
the arbitrariness of $\ell $ can be translated into the arbitrariness of $G$%
), an arbitrary value of $\beta $ and an arbitrary value of $a(s)$, compute
the solution $f(\xi _{i})$, $\xi _{i}=hi$ for $i=1,...,N$ and $h=L/N$ with $%
L $ and $N$ sufficiently large (where $L$ represents the length of the
domain and will be taken large) by computing the integral at the right hand
side of (\ref{inteq}), and check whether $f(L)$ is positive or negative. By
shooting with the parameter $a(s)$ we obtain a solution $f(\xi )$ which is
positive and such that $f(L)$ gets as close as desired to zero. If $L$ is
sufficiently large, such solution is very close to our selfsimilar solution.
After such solution is computed, we numerically evaluate%
\begin{equation*}
G_{out}=\int_{0}^{\infty }\eta ^{s}f(\eta )d\eta .
\end{equation*}%
In general $G_{out}\neq G$ so that the solution constructed is not
consistent with the value of $G$ taken a priori, but by choosing $\beta $
appropriately we can make $G_{out}=G$ therefore finding the similarity
exponent $\beta $. To summarize, our method is a shooting procedure with two
parameters, $a(s)$ and $\beta $, and the two conditions to find these
parameters (or nonlinear eigenvalues) are: 1) the resulting solution is
positive and $f(L)=0$, 2) $G_{out}=G$.

\begin{figure}[h]
\centering\includegraphics[width=.95\textwidth]{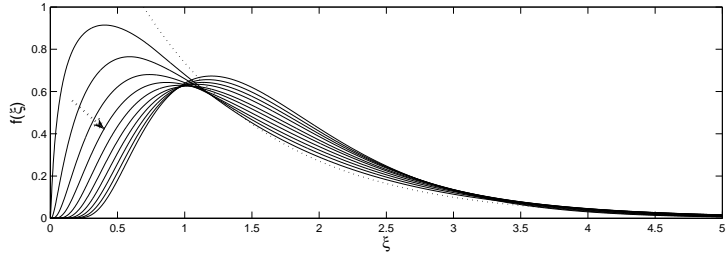}
\caption{{}Similarity solutions for $s=-0.1,-0.2,...,-1$. The arrow
indicates incresing values of $-s$.}
\label{fig3}
\end{figure}

\begin{figure}[h]
\centering\includegraphics[width=.95\textwidth]{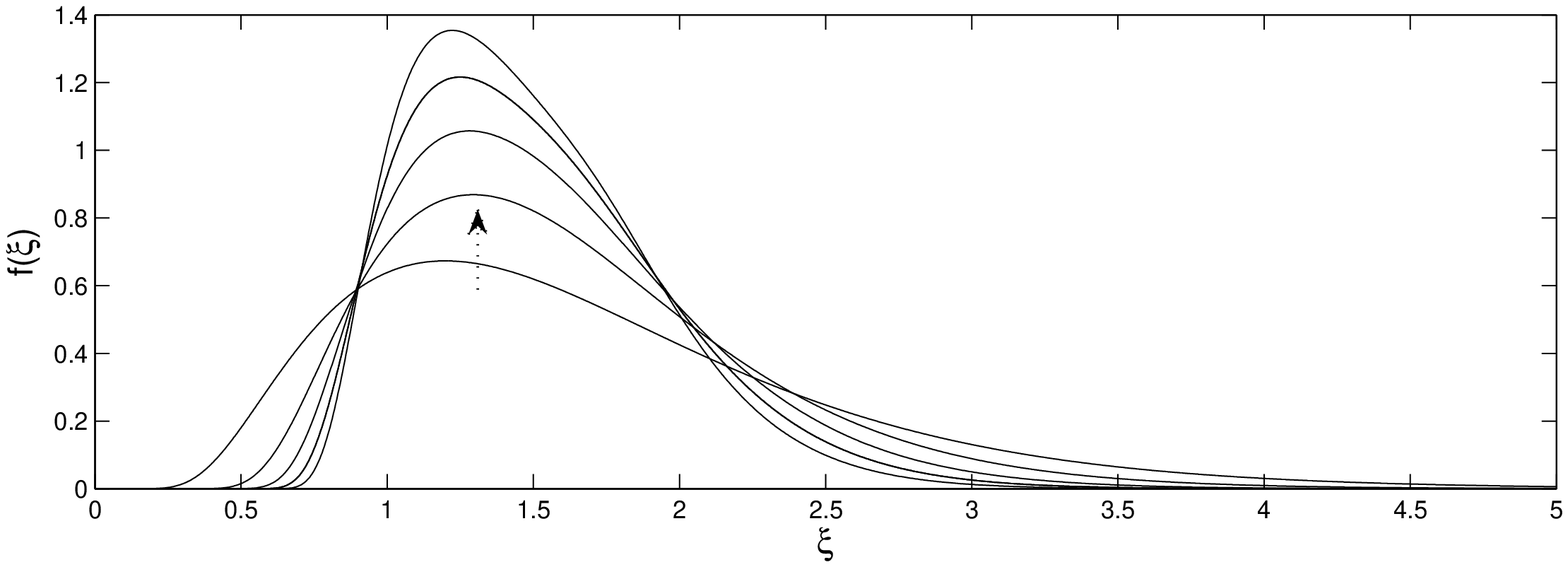}
\caption{{}Similarity solutions for $s=-1,-2,...,-5$. The arrow indicates
incresing values of $-s$.}
\label{fig4}
\end{figure}

\begin{figure}[h]
\centering\includegraphics[width=.95\textwidth]{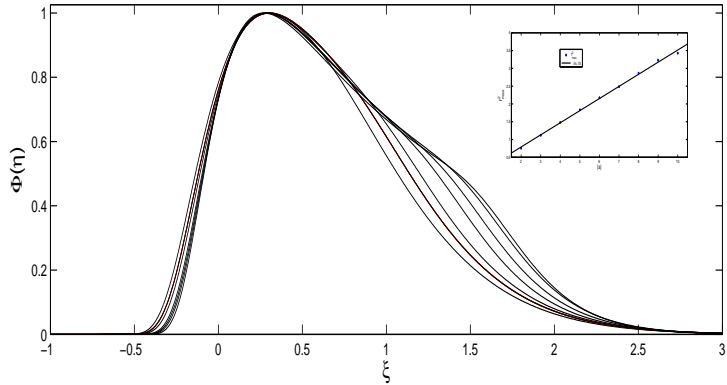}
\caption{{}Rescaled similarity solutions for $-s=4,5,...,10$. Inset: value
of $f_{\max }^{2}$ vs. $\left\vert s\right\vert $ and comparison with a
linear law.}
\label{fig5}
\end{figure}

As it was expected, the numerical values of $\beta $ as a function of $s$
approach the curve%
\begin{equation}
\beta =-\frac{1}{1-2s},  \label{beps}
\end{equation}%
within less than 1\% of relative error. In Figures \ref{fig3}, \ref{fig4} we
represent the similarity solutions for various values of $s$. Notice the
existence of a change in the shape of the similarity solutions as $%
\left\vert s\right\vert $ increases. For small values of $\left\vert
s\right\vert $ the maximum decreases, but eventually, as $\left\vert
s\right\vert $ increases, the maximum starts to grow and the shape of the
similarity solutions can be very well represented by
\begin{equation*}
f(\xi )\simeq \left\vert s\right\vert ^{\frac{1}{2}}\Phi \left( \left\vert
s\right\vert ^{\frac{1}{2}}(\xi -1)\right) ,
\end{equation*}%
for large values of $\left\vert s\right\vert $, as anticipated by (\ref{ssm}%
). In Figure \ref{fig5} we show the collapse of the rescaled (with $\sqrt{%
\left\vert s\right\vert }$) profiles towards a certain function $\Phi $.


\section{Numerical solutions of Smoluchowski equation}

In this section we follow the time evolution of an arbitrary initial
distribution $c_{0}$, numerically treating Smoluchowski's equation as a
differential equation of the form $\partial _{t}c\left( x,t\right) =F\left(
c\left( x,t\right) ,x,t\right) $ with $F$ given as the right hand side of {(}%
\ref{smo}{)}. Our approach has been to adopt a standard predictor-corrector,
fourth order and variable time step integrator. In order to produce the
numerical results, we have used almost the same scheme that was originally
designed by Lee in \cite{Lee}. Other authors have worked out more stable and
sophisticated versions of this algorithm: we point out the recent
contribution of Fibet and Lauren\c{c}ot \cite{Filbet-Lauren=0000E7ot} among
them.

Let $\mathbf{x}$ be a spatially uniform grid ranging from $x_{1}=\delta _{x}$
to $x_{N}=N\delta _{x}$; we will call mass sites or mass bins $x_{k}$
following the way they are commonly referred to in the literature. When the
possible mass numbers are multiples of a minimum $\delta _{x}$,
Smoluchowski's equation reduces to a discrete form:
\begin{equation*}
\partial _{t}c\left( x_{j},t\right) =\frac{1}{2}\delta
_{x}\sum_{l+k=j}K\left( x_{l},x_{k}\right) c\left( x_{l},t\right) c\left(
x_{k},t\right) -c\left( x_{j},t\right) \delta _{x}\sum_{k=1}^{N}K\left(
x_{j},x_{k}\right) c\left( x_{k},t\right) ,
\end{equation*}%
whose right hand side can be easily computed numerically. It is also evident
that this choice cuts off an infinite quantity of mass sites that, sooner or
later, will become dynamically relevant in the system. To avoid this
restriction, a change of variable $x\rightarrow 1/(1+x)$ was used to map the
positive x-axis on the bounded interval $\left( 0,1\right) $, but, as it has
been clearly pointed out in \cite{Filbet-Lauren=0000E7ot}, it is not clear
how to control the distribution of the new mesh points or the mass
distribution among them. See also \cite{Eras-Eyre-Evers} and references
therein for this kind of approach.

In order to determine the cut mass $x_{N}$, our empirical criteria has been
the following: given $T_{f}$ the desired ending time, if the solution has to
reach a selfsimilar regime $c\left( x,t\right) \sim t^{\alpha }\psi \left(
t^{\beta }x\right) $, one can find a proper value for $x$ such that $\left(
T_{f}\right) ^{\alpha }\psi \left( \left( T_{f}\right) ^{\beta }x\right)
\leq tol$, where $tol$ is a numerical parameter indicating the maximum
permitted density of $x_{N}$-massed clusters at the final time; the value of
$\beta $ can be taken coarsely as $\beta \sim -1-2\varepsilon ,$ giving $%
\alpha \sim -2-4\varepsilon -4\varepsilon ^{2}$ and a low accuracy $\psi $
can be computed via a previous low order simulation.

A great advantage of an uniformly distributed bin model is that the
integrodifferential problem is reduced to a $N$-dimensional vector valued
ordinary differential equation. Therefore, standard integration algorithms
can be applied with good performances. A predictor-corrector method quickly
brings an approximation of an implicit scheme, avoiding the heavy workload
that computing $F\left( c\left( t,x\right) ,t,x\right) $ at each step would
impose; it is, moreover, almost possible to guarantee the conservation of
the first moment until the initial mass spreads over the $x$-line,
augmenting significantly the lost mass that have reached the tail. As for
the variable time step method, such an implementation is highly desirable
since the peaks of variation in the distribution of $c$ tend to reduce
quickly as the time passes. It is thus possible to gradually augment $\delta
_{T}$ and still maintain a relative $c$-variation small enough. We refer to
the huge numeric receipts literature for the reader to find further
informations on those classical methods.

To compute the $N$-dimensional vector $F$ we consider all possible binary
interactions $\left\{ i,k\right\} $ between active bins of mass: given a
small numerical threshold $\mu $, we define at each time $t$ the set $%
\mathbf{v}=\left\{ i:c_{i}\left( t\right) \text{\textperiodcentered }%
x_{i}\geq \mu \right\} $. Therefore, in a cycle for $i$ ranging on $\mathbf{v%
}$, we consider $\mathbf{v}_{i}=\left\{ k\in \mathbf{v}:k>i\right\} $ and
for each pair $\left\{ i,k\right\} _{k\in \mathbf{v}_{i}}$
\begin{equation}
F_{i}=F_{i}-\delta _{x}K\left( x_{i},x_{k}\right) c_{i}\left( t\right)
c_{k}\left( t\right) ,\,\,\,\,\,\,\,F_{k}=F_{k}-\delta _{x}K\left(
x_{i},x_{k}\right) c_{i}\left( t\right) c_{k}\left( t\right) ,  \label{eq:F1}
\end{equation}%
and, if $i+k\leq N$,
\begin{equation}
F_{i+k}=F_{i+k}+\delta _{x}K\left( x_{i},x_{k}\right) c_{i}\left( t\right)
c_{k}\left( t\right) .  \label{eq:F2}
\end{equation}%
Notice that we have not included the $\left\{ i,i\right\} $ pair. It is also
necessary to consider it, but it provides only half of the coagulating mass:
\begin{equation}
F_{i}=F_{i}-\delta _{x}K\left( x_{i},x_{i}\right) c_{i}^{2}\left( t\right)
,\,\,\,\,\,\,\,F_{2i}=F_{2i}+\frac{1}{2}\delta _{x}K\left(
x_{i},x_{i}\right) c_{i}^{2}\left( t\right) ,\,\,\,\,\,\,\text{ if}\,\
2i\leq N.  \label{eq:F3}
\end{equation}

A new time step $\Delta t$ is established if the absolute variation between $%
c\left( x,t\right) $ and $c\left( x,t+\Delta t\right) $ is less or equal
than a given tolerance. It is useful to keep track of the evolution of some
relevant moments $M_{\alpha }\left( t+\Delta t\right) $. Since it is
impossible to do it exactly with this finite scheme, we define some
approximated values $m_{\alpha }\left( t+\Delta t\right) $ which resembles $%
M_{\alpha }\left( t+\Delta t\right) $, and, after each new step, we compute:
\begin{equation*}
m_{\alpha }\left( t+\Delta t\right) =\sum_{i=1}^{N}x_{i}^{\alpha
}c_{i}\left( t+\Delta t\right) +\lambda _{\alpha }\left( t+\Delta t\right) ,
\end{equation*}%
where we consider an associated quantity $\lambda _{\alpha }\left( t+\Delta
t\right) $ as the cumulative lost contribution to $m_{\alpha }$. It is
computed in the following way: we consider again all possible binary
interactions $\left\{ i,k\right\} $ between active bins of mass at previous
time $t$ and run a cycle for $i$ ranging on $\mathbf{v}$, but this time we
look only for $\mathbf{v}_{i}^{\infty }=\left\{ k\in \mathbf{v}%
:k>i,k+i>N\right\} $. This set takes into account only the active pairs that
form clusters which exceed the cut mass $x_{N}$. Since $\delta _{x}K\left(
x_{i},x_{k}\right) c_{i}\left( t\right) c_{k}\left( t\right) $ represents
the velocity at which clusters of mass $x_{i+k}$ are being produced and $%
\Delta t$ is the interval of time that has passed, we can approximately
consider that the pair $\left\{ i,k\right\} $ has produced $n_{i,k}\equiv
\Delta t\text{\textperiodcentered }\delta _{x}K\left( x_{i},x_{k}\right)
c_{i}\left( t\right) c_{k}\left( t\right) $ new clusters of mass $x_{i+k}$.
This rough estimate will only be used to compute the lost contribution to $%
m_{\alpha }$:
\begin{equation*}
\lambda _{\alpha }\left( t+\Delta t\right) =\lambda _{\alpha }\left(
t\right) +\sum_{i\in \mathbf{v}}\sum_{k\in \mathbf{v}_{i}^{\infty }}\left(
\left( x_{i}+x_{k}\right) ^{\alpha }-x_{i}^{\alpha }-x_{k}^{\alpha }\right)
n_{i,k}.
\end{equation*}

%\begin{figure}[t]
%\centering%
\begin{figure}[h]
\centering
\includegraphics[width=0.9\textwidth]{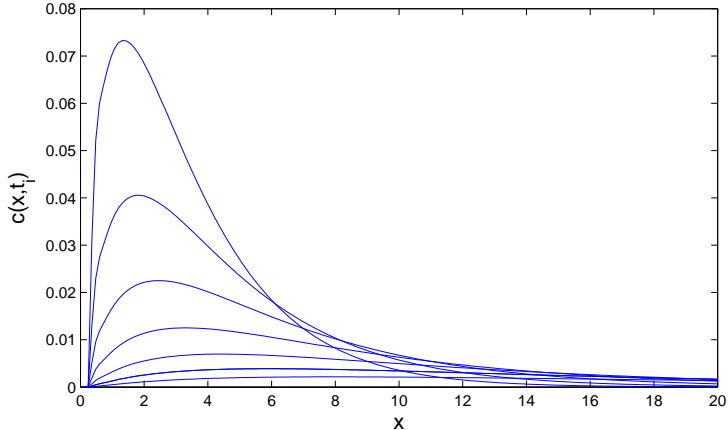}
\caption{Solution of the evolution problem with $\protect\varepsilon =-0.2$
for 7 different times}
\label{evolc1}
\end{figure}
\begin{figure}[h]
\centering
\includegraphics[width=0.9\textwidth]{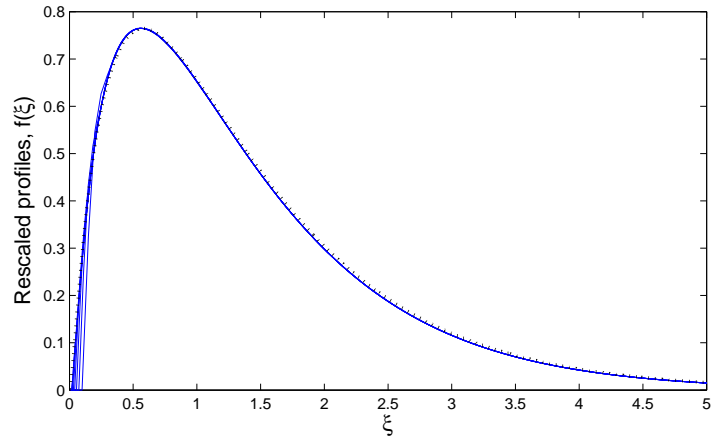}
\caption{Rescaled profiles together with the similarity solution (dotted
line).}
\label{evolc2}
\end{figure}

We remark now that, from the instant when a sufficient mass escapes the
finite coagulating system (infinite mass region), there are three
interactions that are dynamically relevant: finite-finite, infinite-finite
and infinite-infinite mass region coagulation. The former can be numerically
simulated with our scheme while our knowledge of the tail distribution can
only be driven forward via an ansatz (an arbitrary fast decay or a
selfsimilar regime). We preferred nevertheless not to introduce such a tail
into play and make the mass leaving the finite coagulating system completely
stop coagulating. In that resides the need of a $x_{N}$ big enough to
harbour the relevant distribution of $c$ for the solution to go as far as
the self-similar regime. In Figures \ref{evolc1},\ref{evolc2} we present the
result of the evolution of an initial data concentrated close to the origin
and for $s=-0.2 $, together with the rescaled profiles. As we can see, the
convergence towards the selfsimilar solution computed by the procedure
described in the previous section is remarkable.

%
%\begin{subfigure}
%                \includegraphics[width=0.8\textwidth]{evol_eps_p02.eps}
%                \label{fig:evol3}
%        \end{subfigure}%
%\begin{subfigure}
%                \includegraphics[width=0.8\textwidth]{free_sin3.eps}
%                \label{fig:free_sin3}
%        \end{subfigure}
%\caption{Solution of the evolution problem with $\protect\varepsilon =-0.2$
%for 7 different times (top) and the rescaled profiles (bottom) together with
%the similarity solution (dotted line, bottom). }
%\label{evol}
%\end{figure}

\end{document}